\documentclass[10pt,twocolumn,a4paper]{article}
\usepackage[T1]{fontenc}
\usepackage{mathptmx}
\usepackage[english]{babel}
\usepackage[left=1.8cm,right=1.8cm,top=2.2cm,bottom=2.2cm,columnsep=0.7cm]{geometry}
\usepackage[super,sort&compress,comma]{natbib}
\usepackage[version=4]{mhchem}
\usepackage{amsmath,amssymb}
\usepackage{bm}
\usepackage{graphicx}
\usepackage{multirow}
\usepackage{booktabs}
\usepackage{array}
\usepackage{float}
\usepackage[font=small,labelfont=bf,labelsep=period,justification=justified,singlelinecheck=false]{caption}
\usepackage[usenames,dvipsnames]{xcolor}
\usepackage{setspace}
\usepackage{stfloats}\fnbelowfloat
\usepackage{hyperref}
\hypersetup{colorlinks=true,linkcolor=black,citecolor=blue!65!black,urlcolor=blue!65!black}

\addto\captionsenglish{}

\makeatletter
\renewcommand\@biblabel[1]{#1}
\makeatother

\title{\textbf{High-Throughput First-Principles Screening of Unexplored $\bm{\beta}$-MPnX Monolayers for Photocatalysis and Carrier Transport}}
\author{Md. Tanvir Ahmed$^{1,2}$, Saleh Hasan Naqib$^{3}$, and Md. Shahajan Ali$^{1,2,*}$\\[4pt]
{\small\itshape $^{1}$Department of Physics, Pabna University of Science and Technology, Pabna-6600, Bangladesh}\\
{\small\itshape $^{2}$Theoretical and Experimental Advanced Materials Science Lab (TEAMS),}\\
{\small\itshape Pabna University of Science and Technology, Pabna-6600, Bangladesh}\\
{\small\itshape $^{3}$Department of Physics, University of Rajshahi, Rajshahi-6205, Bangladesh}\\
{\small $^{*}$Corresponding author. E-mail: msali@pust.ac.bd}}
\date{}

\DeclareMathSymbol{\planckbar}{\mathord}{AMSb}{"7E}
\begin{document}
\twocolumn[
\begin{@twocolumnfalse}
\maketitle
\vspace{-1.5em}
\begin{abstract}
The $\beta$-MPnX transition-metal pnictide halides remain a largely unexplored family of two-dimensional materials. Here, high-throughput first-principles calculations are employed to investigate 48 $\beta$-MPnX monolayers (M = Ti, Zr, Hf; Pn = P, As, Sb, Bi; X = F, Cl, Br, I), systematically assessing their stability, electronic properties, photocatalytic potential, and carrier-transport characteristics. All 48 compounds exhibit negative formation energies and positive cohesive energies. Combined energetic, mechanical, and dynamical stability assessments identify 12 monolayers for further investigation, of which 10 retain their structural integrity during 5~ps \textit{ab initio} molecular dynamics simulations at 300~K. Hybrid-functional band-structure calculations predict semiconducting behaviour for 11 of the 12 compounds, with band gaps ranging from 0.895 to 1.887~eV, while ZrBiBr remains metallic. Band-edge alignment identifies TiPI, ZrPBr, ZrAsBr, and HfPBr as candidates satisfying the calculated thermodynamic requirements for overall water splitting and the four investigated \ce{CO2}-reduction pathways. Carrier-transport calculations reveal pronounced compositional and directional dependence, with ZrPCl exhibiting an acoustic-phonon-limited electron mobility of $907.25~\mathrm{cm^2\,V^{-1}\,s^{-1}}$ along the $x$-direction and TiAsI showing comparatively high hole mobilities along both in-plane directions. These findings establish composition--property relationships across the $\beta$-MPnX family and identify candidates for further investigation of photocatalytic and charge-transport properties.

\end{abstract}
\vspace{1em}
\end{@twocolumnfalse}
]

\section*{Introduction}

\begin{figure*}[!t]
    \vspace{3mm}
    \centering
    \includegraphics[width=0.80\textwidth]{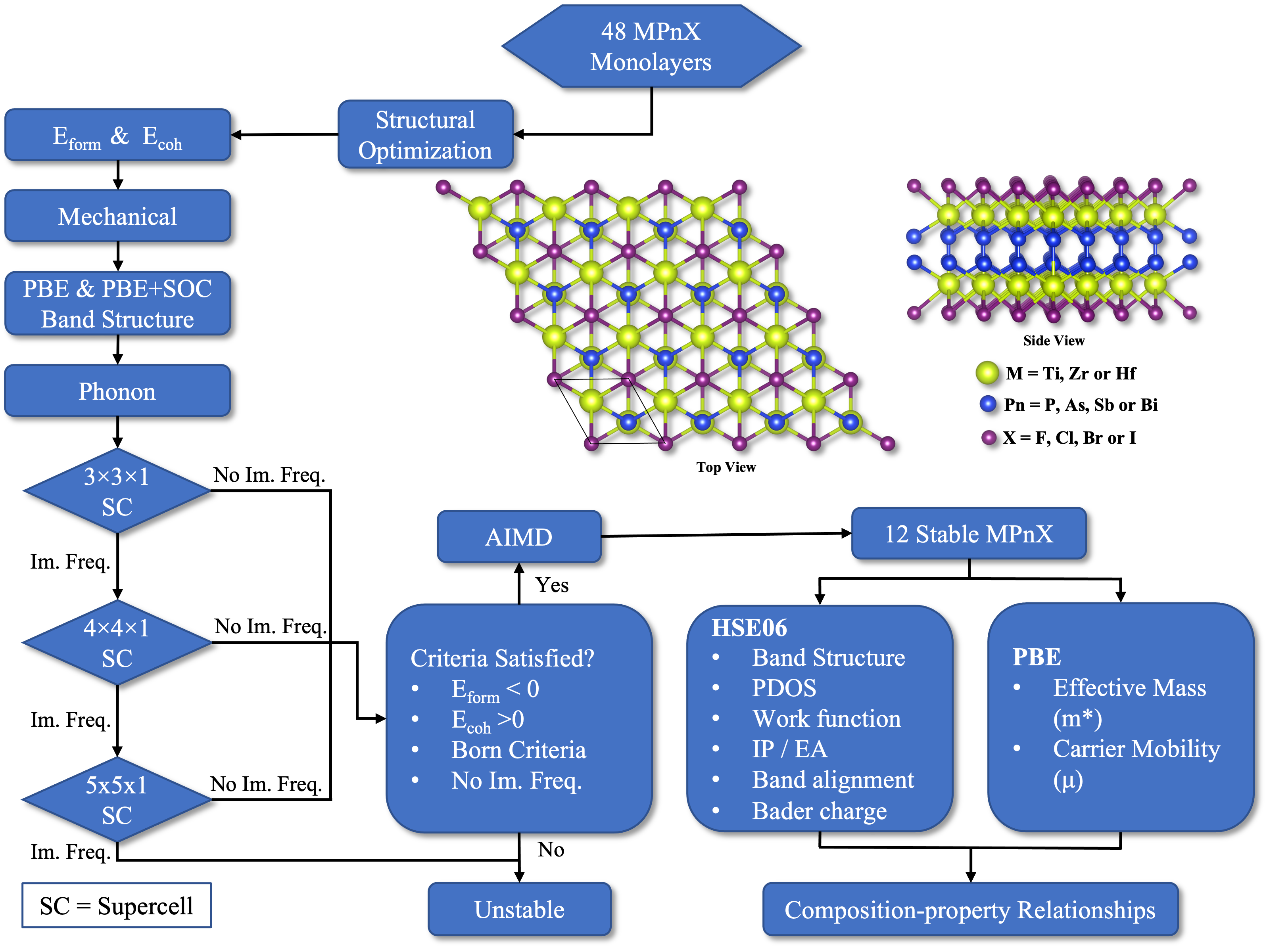}
    \caption{Schematic illustration of the computational workflow for screening 48 $\beta$-MPnX monolayers, including structural optimization, stability assessments, and subsequent investigations of the electronic and carrier-transport properties of the stable compounds.}
    \label{fig:workflow}
    \vspace{-2mm}
\end{figure*}

Two-dimensional (2D) materials have emerged as an important class of materials for next-generation electronic, optoelectronic, and energy-related technologies because of their reduced dimensionality and distinctive electronic, mechanical, optical and transport properties.\cite{Lin2023,Zhuang2025,Ren2025} The discovery of graphene and the subsequent development of transition-metal dichalcogenides, black phosphorus, MXenes and other atomically thin materials have demonstrated the broad potential of low-dimensional systems.\cite{Novoselov2004,Wang2012,Li2014,Naguib2013,Pan2017} Nevertheless, limitations associated with chemical stability, carrier transport, band-gap characteristics and tunability continue to motivate the systematic exploration of new 2D materials with diverse compositions and structural frameworks.\cite{Mounet2018}

Among the diverse families of layered compounds explored as potential 2D materials, transition-metal nitride halides (TMNHs) have attracted considerable attention because of their distinctive layered structures and tunable physical properties.\cite{Hao2020,Xie2023,Kasahara2015} TMNHs generally occur in two distinct layered polymorphs, namely the $\alpha$-type FeOCl structure and the $\beta$-type SmSI structure.\cite{Yamanaka2000,Chen2002} The $\alpha$-type structure is characterized by an orthorhombic layered framework comprising transition-metal--nitrogen layers separated by halogen layers and adopts the $Pmmn$ space group in both bulk and monolayer forms.\cite{Huang2023} In contrast, the $\beta$-type structure is characterized by a double honeycomb M–N layer sandwiched between halogen layers, forming a hexagonal layered framework.\cite{Yun2017} In the bulk $\beta$ phase, the periodic stacking of these layers along the out-of-plane direction gives rise to a rhombohedrally centred $R\bar{3}m$ space group.\cite{Dai2022} Upon isolation of a single layer, the three-dimensional stacking periodicity and associated rhombohedral centring are removed, and the monolayer adopts the primitive $P\bar{3}m1$ space group while retaining the characteristic $\bar{3}m$ point symmetry. The relatively weak interlayer van der Waals interactions in these layered compounds facilitate their exfoliation into atomically thin sheets.\cite{Huang2023,Osanloo2021}

Subsequent studies have progressively established the stability and functional properties of transition-metal nitride halide monolayers. Early investigations of Ti-based compounds demonstrated that $\alpha$-TiNF, $\alpha$-TiNCl, and $\alpha$-TiNBr monolayers are structurally, dynamically, and thermally stable and possess favourable electronic and optoelectronic properties, including suitable band gaps, strong optical absorption, and low exciton binding energies for excitonic solar-cell applications.\cite{Liang2018} Further investigations revealed pronounced anisotropy in electronic and phonon transport in these Ti-based monolayers, with TiNBr exhibiting the best predicted thermoelectric performance.\cite{WangGao2020} In the $\beta$ phase, Yun and Lee demonstrated the feasibility of isolating $\beta$-ZrNCl and $\beta$-HfNCl monolayers through cleavage-energy, phonon-dispersion, and \textit{ab initio} molecular-dynamics calculations, supporting their structural stability and finite-temperature robustness as indirect-gap semiconductors with promising electron-transport and thermoelectric properties.\cite{Yun2017} Subsequent first-principles studies have shown that replacing Zr with Hf substantially modifies the thermal transport properties of transition-metal nitride halides. Notably, HfNCl exhibits approximately fourfold higher lattice thermal conductivity than ZrNCl at 300 K, despite the greater atomic mass of Hf. This anomalous behaviour has been attributed to stronger Hf–N bonding associated with the lanthanide contraction, which broadens the phonon gap, reduces the phase space available for phonon scattering, and suppresses three-phonon scattering.\cite{Yu2020} Beyond metal-site substitution, halogen substitution also offers a route to tuning the properties of these materials. For example, $\beta$-HfNI has been predicted to be dynamically stable, with an indirect band gap of 1.322 eV within GGA, and to exhibit promising thermoelectric performance, with a maximum n-type figure of merit, $ZT$, of 0.91 at 500 K.\cite{Dai2022} Further investigations of $\beta$-HfNCl have explored its mechanical properties, strain-induced indirect-to-direct band-gap transition, and electronic transport, including its potential applications in nanoscale electronic and optoelectronic devices.\cite{Wu2024} Importantly, recent first-principles calculations of $\beta$-ZrNCl and $\beta$-HfNCl have demonstrated that optical-phonon scattering substantially influences room-temperature carrier mobility. Neglecting these contributions in mobility estimates based solely on acoustic deformation-potential theory can therefore lead to a significant overestimation of the carrier mobility.\cite{Liu2025}

\begin{figure*}[!t]
    \centering
    \includegraphics[width=0.90\textwidth]{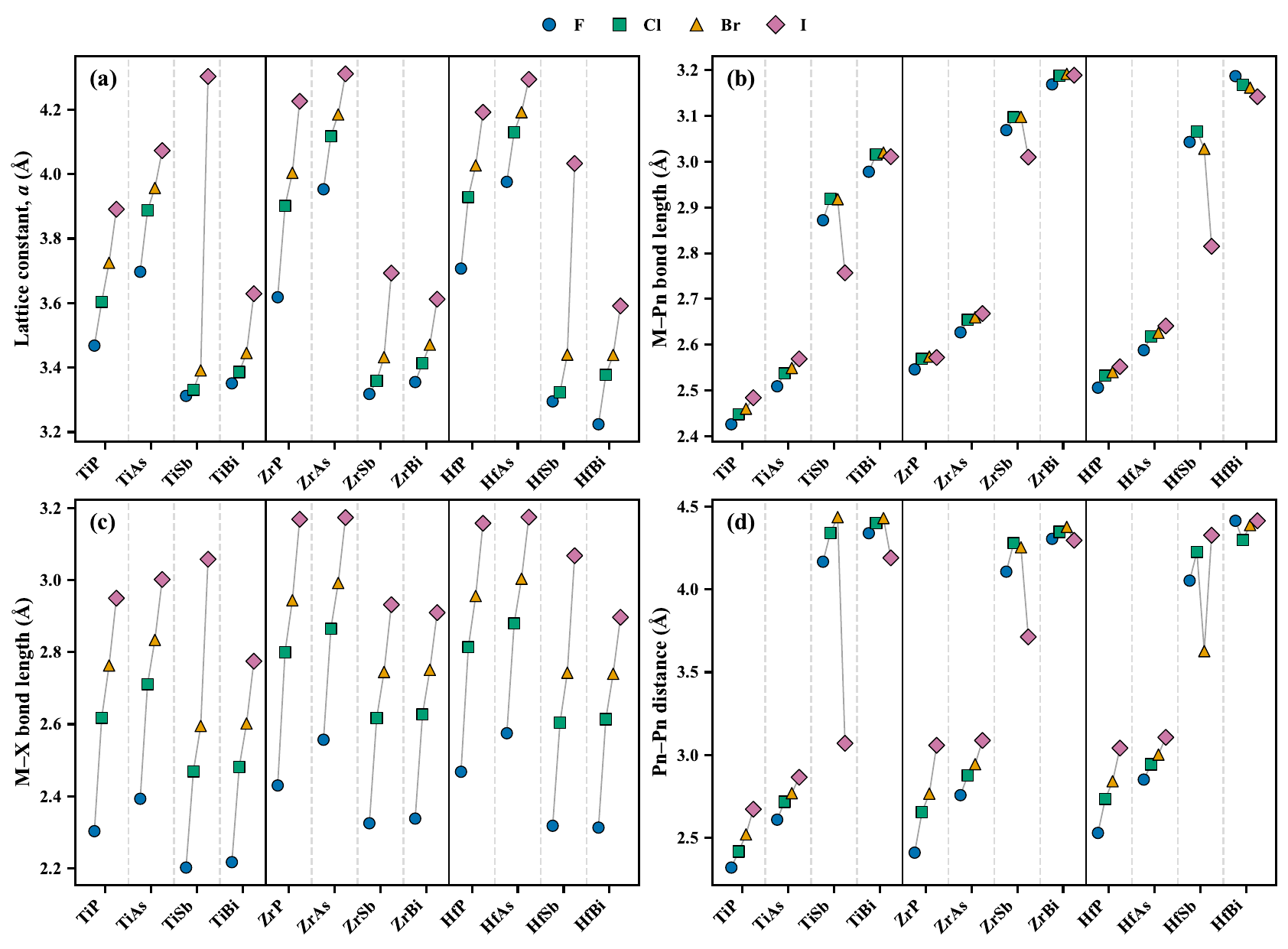}
    \caption{Structural properties of the 48 $\beta$-MPnX monolayers: (a) lattice constant, (b) M--Pn bond length, (c) M--X bond length, and (d) Pn--Pn distance. Different marker shapes and colours distinguish the halogen species (F, Cl, Br, and I).}
    \label{fig:structural}
\end{figure*}

Despite the progress achieved for transition-metal nitride halides, existing studies remain concentrated on a limited number of nitride-based compositions. These studies have addressed structural stability, thermal and electronic transport, thermoelectric and solar-cell applications and, for a few $\alpha$-type and Janus compositions, photocatalytic water splitting, but they have focused almost entirely on nitrogen as the pnictogen.\cite{WangPCCP2023,Huang2023,Xie2023} In contrast, the corresponding transition-metal pnictide halides, in which nitrogen is replaced by P, As, Sb, or Bi, remain largely unexplored. In particular, the broader chemical space arising from the simultaneous variation of the transition metal, pnictogen, and halogen has not been systematically investigated for $\beta$-type monolayers. This gap limits a comprehensive understanding of how chemical composition and bonding characteristics govern the stability and functional properties of this family of materials. To address this gap, we investigate 48 previously unexplored $\beta$-phase transition-metal pnictide halide ($\beta$-MPnX) monolayers, comprising M = Ti, Zr, Hf; Pn = P, As, Sb, Bi; and X = F, Cl, Br, I. Here, we extend the nitride-halide work in three ways. First, the 48 monolayers are screened in one consistent framework, using formation and cohesive energies, Born criteria, phonons and \textit{ab initio} molecular dynamics (AIMD) in sequence. Second, for the monolayers that pass, HSE06 band edges are compared with the water-splitting and four CO$_2$-reduction levels to assess photocatalytic potential. Third, effective masses and deformation-potential mobilities are calculated across the whole family, which shows how the metal, pnictogen and halogen control charge transport. This combination of water-splitting and CO$_2$-reduction criteria distinguishes the present screening from recent first-principles studies of other two-dimensional photocatalysts, such as surface-functionalized MOenes and pentagonal Si-based monolayers, which have focused mainly on water splitting.\cite{WangMOene2026,YangSi2025} This study establishes composition--property relationships across the $\beta$-MPnX family and identifies promising candidates for further theoretical and experimental investigation.

%%%%%%%%%%%%%%%%%%%%%%%%%%%%%%%%%%%%%%%
%%%%%%%%%%%%%%Method%%%%%%%%%%%%%%%%%%%
%%%%%%%%%%%%%%%%%%%%%%%%%%%%%%%%%%%%%%%
\section*{Computational Methods}
First-principles calculations were performed within the framework of density functional theory (DFT) using the Vienna \textit{ab initio} Simulation Package (VASP).\cite{KresseFurthmuller1996} The projector augmented-wave (PAW) method was employed to describe the electron--ion interactions,\cite{KresseJoubert1999} while the exchange--correlation energy was treated using the Perdew--Burke--Ernzerhof (PBE) generalized gradient approximation (GGA).\cite{Perdew1996} The plane-wave energy cutoff was set to 1.5 times the maximum recommended cutoff among the constituent PAW potentials. The Brillouin zone was sampled using $\Gamma$-centred k-point meshes with a reciprocal-space separation of $0.03~\text{\AA}^{-1}$. All structures were fully relaxed until the total-energy and residual-force convergence criteria reached $10^{-8}$~eV and $0.001~\mathrm{eV}\,\text{\AA}^{-1}$, respectively.

The energetic stability of the optimized monolayers was evaluated using formation and cohesive energies. Formation energies were calculated relative to the corresponding elemental reference states, with the reference structures selected from the Materials Project\cite{Jain2013MaterialsProject} database based on zero energy above the convex hull. Cohesive energies were determined relative to the energies of the constituent isolated atoms. Mechanical stability was investigated using the energy--strain method.\cite{Zhao2007ElasticConstants} In-plane strains ranging from $-1.5\%$ to $+1.5\%$ in increments of $0.5\%$ were applied to the optimized structures. The resulting total energies were fitted using second-order polynomials, and the in-plane elastic constants, $C_{ij}$, were obtained from the corresponding second derivatives of the fitted energy–strain curves, normalized by the equilibrium area of the monolayer. Mechanical stability was assessed using the Born stability criteria.\cite{Wang2022HighThroughput2D}

\begin{figure*}[!t]
\centering
\includegraphics[width=0.90\textwidth]{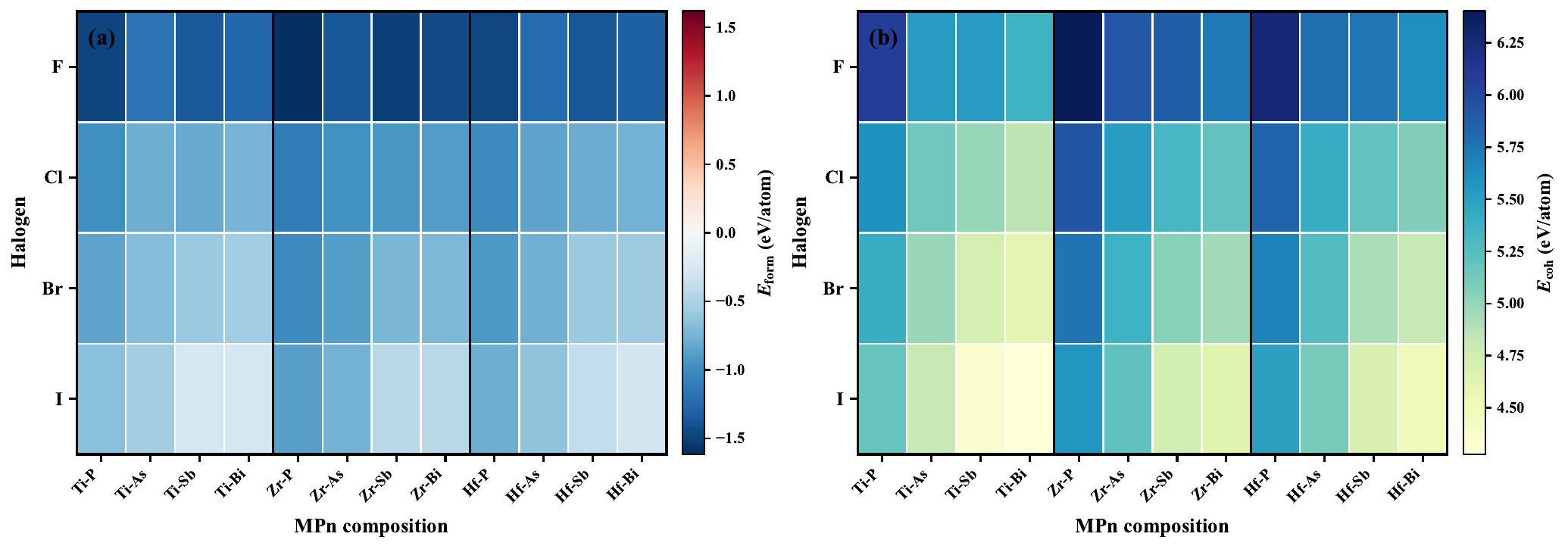}
\caption{Formation and cohesive energies of the 48 $\beta$-MPnX monolayers across the investigated chemical space. (a) Formation energy, $E_{\mathrm{form}}$, and (b) cohesive energy, $E_{\mathrm{coh}}$, shown as colour-coded heatmaps. The formation-energy scale is centred at zero to distinguish negative and positive formation energies, while the cohesive-energy scale represents the magnitude of the cohesive energy in eV/atom.}
\label{fig:energetics}
\end{figure*}

The electronic structures of all monolayers were calculated using PBE, both without and with spin--orbit coupling (SOC). Phonon dispersion relations were calculated using the finite-displacement method implemented in the Phonopy package.\cite{TogoTanaka2015Phonon} Candidates satisfying the energetic, mechanical, and dynamical stability criteria were subsequently examined using \textit{ab initio} molecular dynamics (AIMD) simulations.\cite{KresseHafner1993AIMD} The AIMD simulations were performed in the canonical (NVT) ensemble using a Nosé–Hoover thermostat at 300~K for 5~ps, with a time step of 1~fs. For selected candidates, the simulations were extended to 10~ps under the same conditions.

For the selected stable monolayers, the electronic structures and band-edge characteristics were further investigated using the HSE06 hybrid functional.\cite{Heyd2003Hybrid} The HSE06 band structures and density of states (DOS) were obtained from the same self-consistent calculation, with 25\% exact exchange and a screening parameter of 0.2~\AA$^{-1}$,\cite{Krukau2006} in which the uniform Brillouin-zone mesh determined the charge density and DOS, while additional zero-weighted k-points along the high-symmetry path provided the band dispersion. The optical absorption coefficients were calculated from the HSE06 frequency-dependent dielectric function within the independent-particle approximation, using 72 bands and a complex shift of 0.1~eV, and averaged over the two in-plane directions. Bader charge analysis was also performed using HSE06 to evaluate the charge distribution among the constituent atoms.\cite{Henkelman2005Bader} Carrier effective masses were calculated from the curvature of the corresponding band extrema, and carrier mobilities were estimated using Bardeen--Shockley deformation-potential theory\cite{BardeenShockley1950} within the PBE framework. A denser k-point mesh, corresponding to a reciprocal-space separation of $0.015~\text{\AA}^{-1}$, was employed to determine the band-edge positions for the effective-mass and carrier-mobility calculations.

\section*{Results and discussion}
A systematic screening strategy was employed to identify stable monolayers among the 48 $\beta$-MPnX compounds, where M = Ti, Zr, or Hf; Pn = P, As, Sb, or Bi; and X = F, Cl, Br, or I. The screening involved structural optimization, energetic and mechanical stability assessments, and phonon dispersion calculations, followed by AIMD simulations for candidates satisfying the initial stability criteria. This procedure identified 12 stable monolayers for further investigation of their electronic and carrier-transport properties. The computational workflow and crystal structures of the investigated monolayers are illustrated in  Fig.~\ref{fig:workflow}.

The structural parameters of the 48 $\beta$-MPnX monolayers, including the lattice constants and interatomic distances, are summarized in Fig.~\ref{fig:structural} and Table S1. For all three metal families (M = Ti, Zr, and Hf), the lattice constant generally increases from F to I within each M–Pn series. For example, a increases from 3.468 to 3.891~{\AA} for TiPX, from 3.618 to 4.226~{\AA} for ZrPX, and from 3.707 to 4.192~{\AA} for HfPX as X changes from F to I. This trend is observed across all four pnictogen series (Pn = P, As, Sb, and Bi) for each metal family. Across the pnictogen series, the As-based compounds generally exhibit larger lattice constants than their P-based counterparts, whereas the Sb and Bi-based compounds tend to have smaller lattice constants, although some deviations occur, particularly among the iodides in the Sb-based series.

The M–X bond lengths generally increase from F to I (Fig.~\ref{fig:structural}c), consistent with the increasing atomic size of the halogen species. Across the 48 monolayers, these bond lengths range from 2.202 to 3.175~{\AA}. The M–Pn bond lengths also depend on pnictogen identity (Fig.~\ref{fig:structural}b), with the P and As-based compounds generally exhibiting shorter bonds than their Sb and Bi-based counterparts. These distances span 2.426–3.192~{\AA}, although their variation with halogen substitution depends on the specific metal--pnictogen combination. The Pn–Pn distances also exhibit a distinct compositional dependence (Fig.~\ref{fig:structural}d), ranging from 2.321 to 4.436~{\AA}. The P and As-based monolayers generally exhibit shorter Pn–Pn distances, whereas their Sb and Bi-based counterparts show larger separations. The larger Pn–Pn separations are generally accompanied by greater monolayer thicknesses in the Sb and Bi-based compounds than in the P and As-based compounds, as reported in Table S1. Nevertheless, variations within the Sb and Bi-based series indicate that these structural trends are not strictly monotonic across all metal and halogen combinations.

The energetic stability of the 48 $\beta$-MPnX monolayers was further assessed using their formation and cohesive energies, as presented in Fig.~\ref{fig:energetics}. The formation energy per atom was calculated as\cite{He2019DFT}

\begin{equation}
E_{\mathrm{form}} =
\frac{
E_{\mathrm{tot}}(\mathrm{MPnX})
- N_M \mu_M
- N_{\mathrm{Pn}} \mu_{\mathrm{Pn}}
- N_X \mu_X
}{N}
\label{eq:formation_energy}
\end{equation}

where $E_{\mathrm{tot}}(\mathrm{MPnX})$ is the total energy of the monolayer unit cell, $N_i$ is the number of atoms of species $i$, $N$ is the total number of atoms in the unit cell, and $\mu_i$ is the energy per atom of the corresponding elemental reference phase. Negative formation energies indicate that the compounds are energetically favourable relative to their constituent elements in the selected reference states. The cohesive energy per atom was calculated as

\begin{equation}
E_{\mathrm{coh}} =
\frac{
N_M E_M^{\mathrm{atom}}
+ N_{\mathrm{Pn}} E_{\mathrm{Pn}}^{\mathrm{atom}}
+ N_X E_X^{\mathrm{atom}}
- E_{\mathrm{tot}}(\mathrm{MPnX})
}{N}
\label{eq:cohesive_energy}
\end{equation}

where $E_i^{\mathrm{atom}}$ denotes the energy of an isolated atom of species $i$. Positive cohesive energies indicate that the monolayers are energetically bound relative to their constituent isolated atoms.

\begin{figure*}[!t]
\centering
\includegraphics[width=0.90\textwidth]{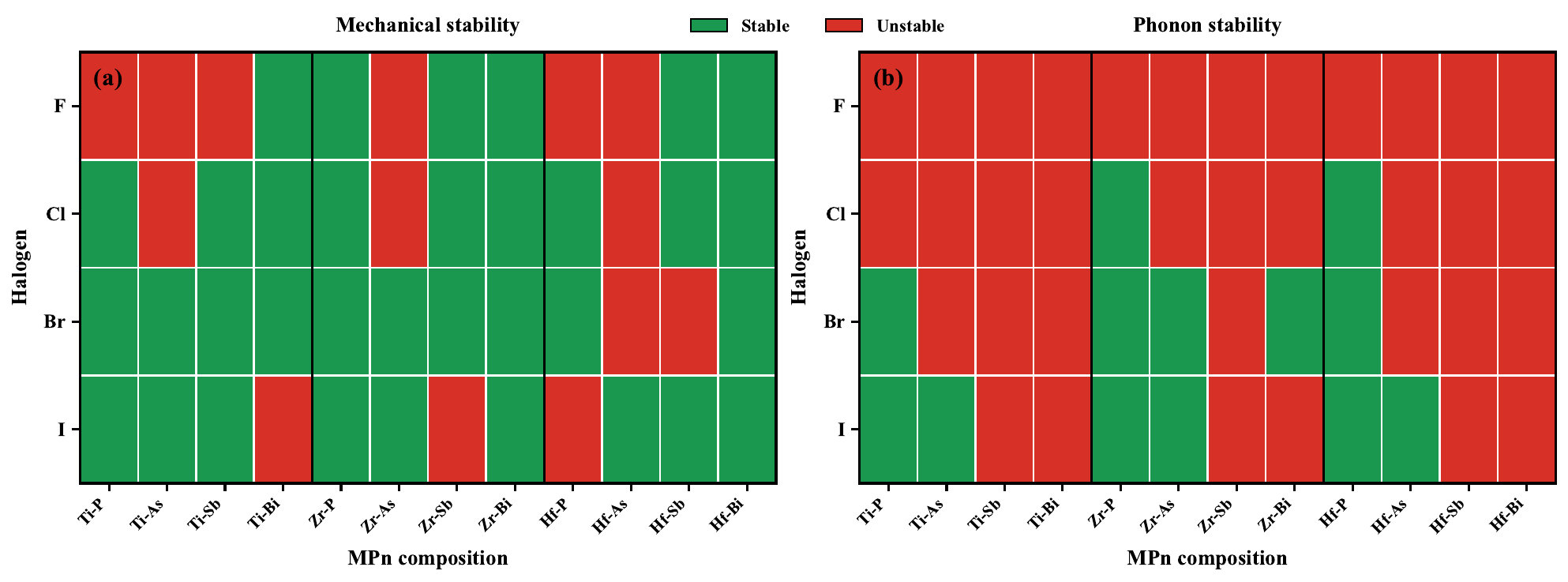}
\caption{Mechanical and dynamical stability of the 48 $\beta$-MPnX monolayers. (a) Mechanical stability evaluated using the Born stability criteria and (b) phonon stability determined from phonon dispersion calculations. Green and red cells indicate structures satisfying and not satisfying the corresponding stability criterion, respectively.}
\label{fig:stability}
\end{figure*}

As shown in Fig.~\ref{fig:energetics}(a,b), all 48 monolayers exhibit negative formation energies and positive cohesive energies. The formation energies range from $-$1.617 $\mathrm{eV/atom}$ for ZrPF to $-$0.282 $\mathrm{eV/atom}$ for TiSbI, while the cohesive energies range from 4.277 $\mathrm{eV/atom}$ for TiBiI to 6.401 $\mathrm{eV/atom}$ for ZrPF. These results indicate favourable formation relative to the selected elemental reference states and substantial energetic binding relative to isolated constituent atoms.

The heatmaps in Fig.~\ref{fig:energetics} reveal a systematic dependence of both energies on halogen substitution. Across all 12 metal–pnictogen combinations, the formation energy becomes progressively less negative from F to I, accompanied by a decrease in cohesive energy. This trend is consistent across the Ti, Zr, and Hf-based families, indicating that halogen substitution systematically influences the energetics of these monolayers. The formation energies also depend on metal identity, with Zr-based compounds generally exhibiting more negative values than their Ti and Hf-based counterparts. Among the 48 compounds, ZrPF exhibits both the most negative formation energy ($-$1.617 $\mathrm{eV/atom}$) and the highest cohesive energy (6.401 $\mathrm{eV/atom}$).

The energetic analysis provides an initial assessment of the favourability of the 48 $\beta$-MPnX monolayers relative to their constituent elemental reference states. However, energetic favourability alone does not establish whether these structures can withstand mechanical deformation or remain dynamically stable. Therefore, their mechanical and dynamical stability was further investigated using the Born stability criteria and phonon dispersion calculations, respectively, as summarized in Fig.~\ref{fig:stability}.

The mechanical stability of the 48 $\beta$-MPnX monolayers was evaluated from their in-plane elastic response using the Born stability criteria. For a hexagonal monolayer, the elastic stiffness relation can be expressed as\cite{Zhao2007ElasticConstants,Wang2022HighThroughput2D,Mazdziarz2019Comment}

\begin{equation}
\begin{pmatrix}
\sigma_x \\
\sigma_y \\
\tau_{xy}
\end{pmatrix}
=
\begin{pmatrix}
C_{11} & C_{12} & 0 \\
C_{12} & C_{11} & 0 \\
0 & 0 & C_{66}
\end{pmatrix}
\begin{pmatrix}
\varepsilon_x \\
\varepsilon_y \\
\gamma_{xy}
\end{pmatrix}
\label{eq:elastic_stiffness}
\end{equation}

Here, $C_{11}$, $C_{12}$, and $C_{66}$ represent the longitudinal, transverse coupling, and shear elastic constants, respectively. The elastic energy variation associated with small in-plane normal strains is given by

\begin{equation}
\Delta E =
\frac{A_0}{2}
\left(
C_{11}\varepsilon_x^2
+ C_{11}\varepsilon_y^2
+ 2C_{12}\varepsilon_x\varepsilon_y
\right)
\label{eq:elastic_energy}
\end{equation}
    
where $A_0$ is the equilibrium in-plane area of the unit cell. The longitudinal elastic constant was obtained from the curvature of the strain–energy curve under uniaxial strain,

\begin{equation}
C_{11} =
\frac{1}{A_0}
\frac{\partial^2 E}{\partial \varepsilon_x^2}
\label{eq:C11}
\end{equation}
    
while $C_{12}$ was determined from the equal-biaxial strain
($\varepsilon_x = \varepsilon_y = \varepsilon$) response according to

\begin{equation}
C_{12} =
\frac{1}{2A_0}
\frac{\partial^2 E}{\partial \varepsilon^2}
- C_{11}
\label{eq:C12}
\end{equation}
 
For the hexagonal systems considered here, the shear elastic constant is related to the two independent normal elastic constants by
\begin{equation}
C_{66} = \frac{C_{11}-C_{12}}{2}
\label{eq:C66}
\end{equation}

The calculated $C_{11}$, $C_{12}$, and $C_{66}$ values for all 48 monolayers are provided in Table S2. Mechanical stability was assessed using the Born criteria,\cite{Wang2022HighThroughput2D,Mazdziarz2019Comment}

\begin{equation}
C_{11} > 0,
\qquad
C_{11} > \left|C_{12}\right|
\label{eq:born_criteria}
\end{equation}

As summarized in Fig.~\ref{fig:stability}(a), 34 of the 48 monolayers satisfy both Born stability conditions, whereas 14 are mechanically unstable. The calculated in-plane elastic constants and corresponding mechanical stability classifications for all 48 monolayers are provided in Table S2.

The dynamical stability of the monolayers was further assessed using phonon dispersion calculations. As shown in Fig.~\ref{fig:stability}(b), 13 compounds exhibit phonon stability, while the remaining 35 are classified as phonon-unstable. The corresponding phonon dispersion relations and phonon density of states (DOS) for all 48 monolayers are provided in Figs. S1–S6. Comparison of the mechanical and phonon stability results reveals that HfPI is the only compound that exhibits phonon stability despite failing the Born criteria.

\begin{figure*}[!t]
\centering
\includegraphics[width=0.90\textwidth]{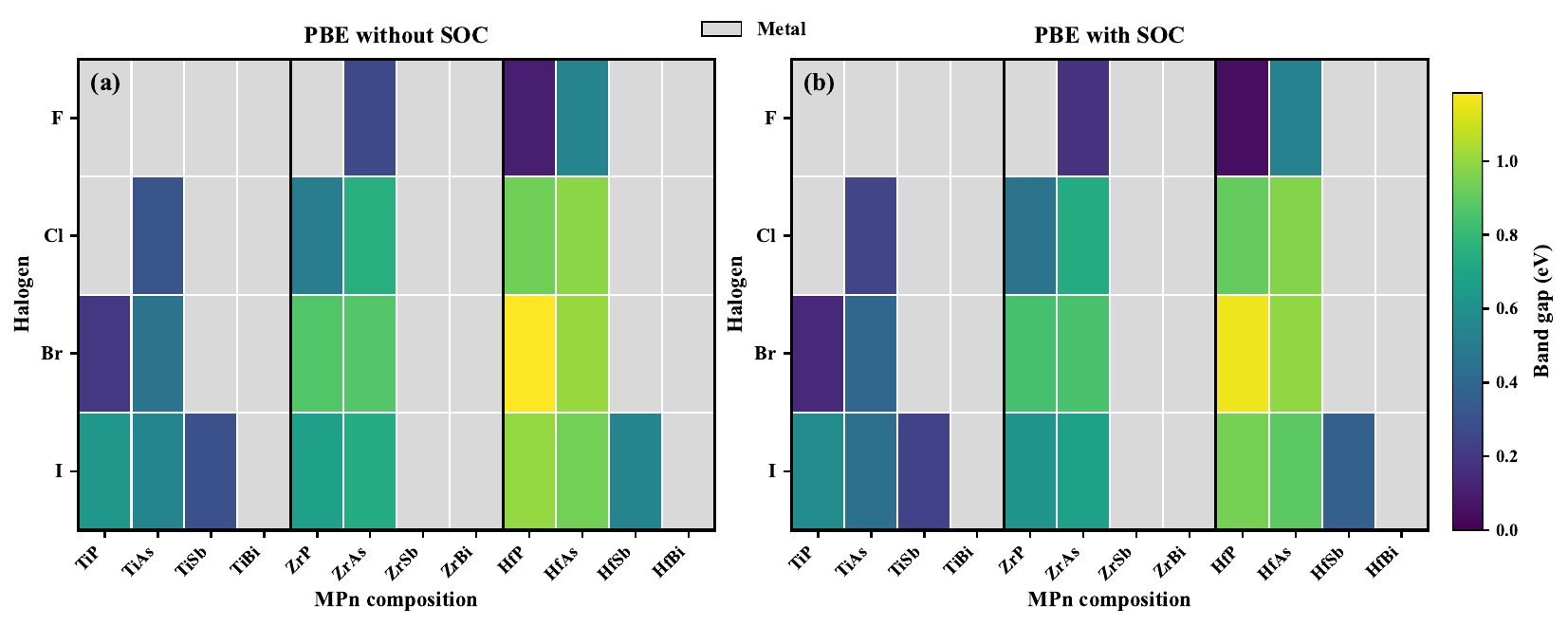}
\caption{PBE band gaps of the 48 $\beta$-MPnX monolayers across the investigated chemical space. (a) Calculated band gaps without SOC and (b) with SOC. The band gaps are represented by a continuous colour scale, while metallic systems are indicated in grey. The common colour scale is referenced to the positive band-gap values, with zero corresponding to the boundary between nonmetallic and metallic behaviour.}
\label{fig:pbe_bandgaps}
\end{figure*}
\begin{figure*}[!b]
    \centering
    \vspace{3mm}
    \includegraphics[width=0.90\textwidth]{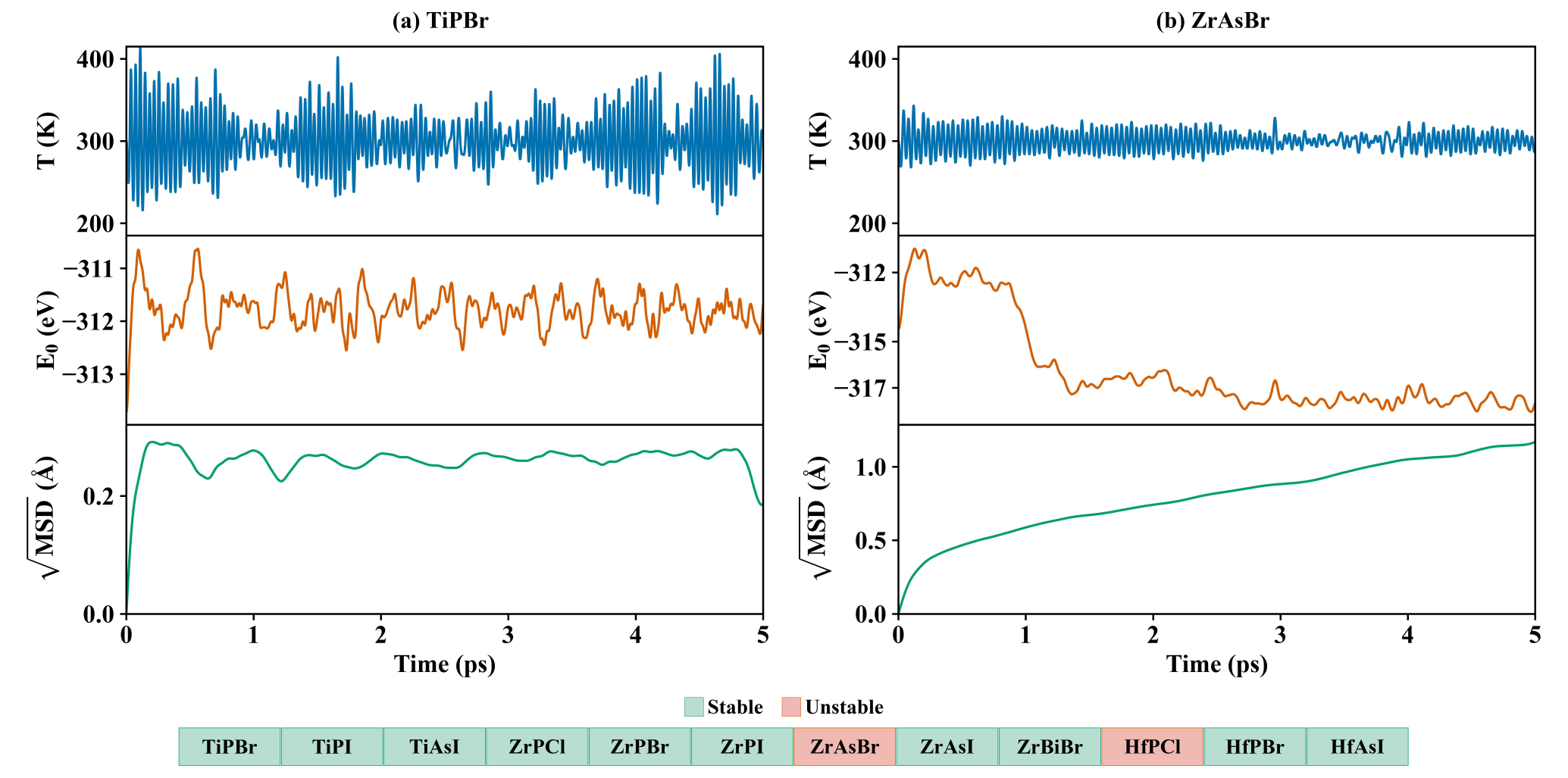}
    \caption{AIMD simulations at 300~K for representative thermally stable and unstable $\beta$-MPnX monolayers: (a) TiPBr and (b) ZrAsBr, over a simulation period of 5~ps. The panels show the evolution of temperature $T$, total energy $E_0$, and $\sqrt{\mathrm{MSD}}$ during the simulations. The stability strip summarizes the AIMD behaviour of all 12 compounds, with green and red indicating stable and unstable trajectories, respectively.}
    \label{fig:aimd}
\end{figure*}

The electronic band structures of all 48 monolayers were calculated using PBE, both without and with SOC. The corresponding band gaps are summarized in Fig.~\ref{fig:pbe_bandgaps}(a,b), while the calculated electronic band structures and densities of states (DOS) are presented in Figs. S7–S18. Inclusion of SOC generally produces relatively small changes in the calculated band gaps, although larger reductions occur in TiAsI and HfSbI, with decreases of approximately 0.11 and 0.18~eV, respectively. Despite these changes, the electronic classification of all 48 compounds remains unchanged, with 26 classified as metallic and the remaining 22 exhibiting positive band gaps. Notably, 25 of the 26 metallic compounds also exhibit phonon instabilities. ZrBiBr is the only metallic compound that satisfies both the mechanical and dynamical stability criteria. Thus, phonon instability is prevalent among the metallic compounds identified in this screening. The pnictogen has the strongest influence on this behaviour. All 12 Bi-based and 10 of the 12 Sb-based compounds are metallic, compared with only four of the 24 P and As-based compounds. Because the valence-band edge contains pnictogen $p$ states, the higher-lying $p$ levels of the less electronegative Sb and Bi atoms raise the VBM and close the gap, and the resulting metallic character is accompanied by dynamical instability. The fluorides form a separate case. Although they have the most favourable formation energies, all 12 are dynamically unstable, which shows that favourable energetics alone do not ensure dynamical stability.

Considering the combined energetic, mechanical, and dynamical stability assessments, 12 monolayers were selected for further investigation: TiPBr, TiPI, TiAsI, ZrPCl, ZrPBr, ZrPI, ZrAsBr, ZrAsI, ZrBiBr, HfPCl, HfPBr, and HfAsI. These compounds exhibit negative formation energies and positive cohesive energies, satisfy the Born stability criteria, and show no imaginary phonon frequencies in the final phonon calculations.

\begin{table*}[!t]
\centering
\caption{HSE06-derived Bader charge transfer and electronic and surface properties of the screened $\beta$-MPnX monolayers. The table lists the charge transfer at the transition-metal, pnictogen and halogen sites, together with the HSE06 band gap ($E_g$), work function ($\Phi$), ionization potential (IP) and electron affinity (EA).}
\label{tab:electronic_properties}

\renewcommand{\arraystretch}{1.25}

\begin{tabular}{
>{\centering\arraybackslash}m{1.5cm}
*{7}{>{\centering\arraybackslash}m{1.5cm}}
}
\hline
Compound
& $\Delta q(M)$
& $\Delta q(\mathrm{Pn})$
& $\Delta q(X)$
& $E_g$ (eV)
& $\Phi$ (eV)
& IP (eV)
& EA (eV) \\
\hline
TiPBr  & 1.561 & -0.901 & -0.655 & 0.895  & 4.875 & 5.128 & 4.233 \\
TiPI   & 1.553 & -0.972 & -0.578 & 1.526  & 5.092 & 5.322 & 3.796 \\
TiAsI  & 1.507 & -0.908 & -0.596 & 1.404  & 5.003 & 5.233 & 3.829 \\
ZrPCl  & 1.864 & -1.104 & -0.756 & 1.037  & 4.741 & 4.985 & 3.948 \\
ZrPBr  & 1.875 & -1.168 & -0.703 & 1.546  & 5.208 & 5.379 & 3.833 \\
ZrPI   & 1.866 & -1.261 & -0.602 & 1.256  & 4.961 & 5.173 & 3.916 \\
ZrAsBr & 1.815 & -1.098 & -0.714 & 1.555  & 5.163 & 5.393 & 3.838 \\
ZrAsI  & 1.793 & -1.163 & -0.627 & 1.341  & 4.876 & 5.146 & 3.805 \\
ZrBiBr & 1.160 & -0.470 & -0.690 & -1.190 & 3.804 & ---   & ---   \\
HfPCl  & 1.907 & -1.164 & -0.742 & 1.540  & 4.984 & 5.154 & 3.614 \\
HfPBr  & 1.900 & -1.208 & -0.689 & 1.887  & 5.250 & 5.446 & 3.559 \\
HfAsI  & 1.792 & -1.171 & -0.618 & 1.500  & 4.989 & 5.229 & 3.728 \\
\hline
\end{tabular}

\end{table*}

To examine the finite-temperature structural stability of these stable monolayers, AIMD simulations were performed at 300~K over a period of 5~ps. Fig.~\ref{fig:aimd} presents the evolution of temperature, total energy, and the square root of the mean-square displacement ($\sqrt{\mathrm{MSD}}$) for two representative compounds, TiPBr and ZrAsBr. TiPBr exhibits bounded fluctuations in total energy and $\sqrt{\mathrm{MSD}}$ throughout the simulation, indicating that its structure remains intact at 300~K. In contrast, ZrAsBr exhibits a pronounced decrease in total energy and a continuous increase in $\sqrt{\mathrm{MSD}}$, suggesting substantial atomic rearrangement and structural instability under the simulated conditions. The corresponding AIMD trajectories for all 12 stable monolayers are presented in Fig. S19. Among these compounds, ZrAsBr and HfPCl exhibit unstable behaviour during the 5~ps simulations, whereas the other 10 retain their structural integrity at 300~K. These results characterize the thermal behaviour of the 12 monolayers under the conditions investigated. The observed instability of individual compounds at 300~K does not, by itself, establish their instability at lower temperatures. Because a 5~ps trajectory tests only short-time structural stability, the simulations were extended to 10~ps for TiPI, ZrPBr and HfPBr, the band-edge candidates identified below that remain stable within 5~ps. The fourth candidate, ZrAsBr, is already unstable within this time. As shown in Fig.~S20, all three retain their structural integrity, with bounded fluctuations in temperature and total energy throughout the extended runs. However, picosecond-scale AIMD cannot capture slower processes, such as gradual reconstruction or decomposition, so these results indicate stability on the simulated time scale rather than prove long-term stability.

Although $\beta$-MPnX compounds have not yet been reported experimentally, several observations support their feasibility. Their nitride analogues, $\beta$-MNX, have been synthesized as layered bulk crystals, by high-pressure methods for the $\beta$ polymorphs\cite{Yamanaka2000,Chen2002} and by chemical vapour transport for ZrNCl, which has been mechanically exfoliated into thin flakes.\cite{Saito2015} The stable $\beta$-MPnX monolayers identified here show negative formation energies, mechanical and dynamical stability, and structural integrity at 300~K, suggesting that analogous solid-state or vapour-transport growth of layered bulk crystals followed by mechanical or liquid-phase exfoliation is a plausible route, although experimental confirmation is still required.

\begin{figure*}[!b]
    \centering
    \includegraphics[width=0.90\textwidth]{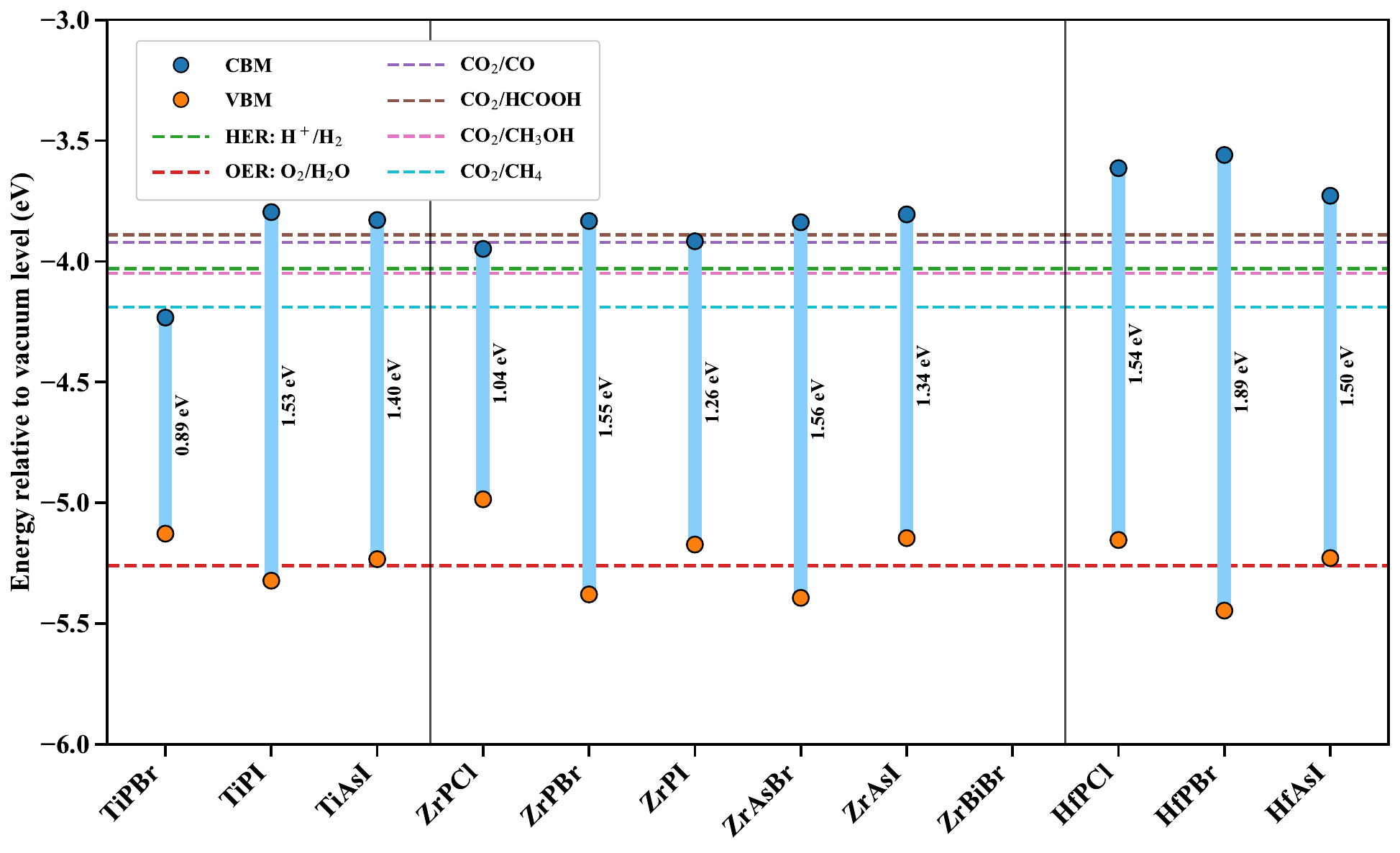}
    \caption{Vacuum-referenced band-edge alignment of the $\beta$-MPnX monolayers calculated using the HSE06 functional. The light-blue vertical segments represent the band gaps, with the corresponding values indicated. The dashed horizontal lines represent the redox levels for water splitting and $\ce{CO2}$ reduction at pH~$=7$ and 25~$^\circ$C. Metallic ZrBiBr is shown without band-edge values.}
    \label{fig:photocatalysis}
\end{figure*}

The electronic structures of the 12 selected $\beta$-MPnX monolayers were further investigated using the HSE06 hybrid functional to obtain a more accurate description of their band gaps and electronic states. The calculated band structures and corresponding DOS are presented in Figs. S21 and S22, while the resulting electronic and surface properties are summarized in Table~\ref{tab:electronic_properties}. The HSE06 calculations predict semiconducting behaviour for 11 compounds, with band gaps ranging from 0.895~eV for TiPBr to 1.887~eV for HfPBr, whereas ZrBiBr retains its metallic character. The projected density of states reveals a consistent orbital contribution near the band edges across the semiconducting compounds. The conduction-band-edge states are dominated by transition-metal $d$ orbitals, whereas the valence-band-edge states primarily originate from transition-metal $d$ and pnictogen $p$ orbitals. The band gaps also depend systematically on the transition metal. Replacing Zr with Hf increases the HSE06 gap from 1.04 to 1.54~eV between ZrPCl and HfPCl, from 1.55 to 1.89~eV between ZrPBr and HfPBr and from 1.34 to 1.50~eV between ZrAsI and HfAsI. Because the conduction-band edge is dominated by metal $d$ states, this increase reflects a higher-lying CBM in the Hf compounds, consistent with their smaller electron affinities. Owing to the lanthanide contraction, Zr and Hf have nearly identical atomic radii, so the difference arises from the higher energy of the Hf 5$d$ states rather than from structural differences. The same trend occurs in the parent nitride halides, where the HSE06 gap of $\beta$-HfNCl (3.37~eV) exceeds that of $\beta$-ZrNCl (2.93~eV).\cite{Yun2017} The Ti compounds show no consistent trend relative to their Zr counterparts.

Among the semiconductors, the calculated work functions range from 4.741~eV for ZrPCl to 5.250~eV for HfPBr. The ionization potentials (IP) of the 11 semiconductors range from 4.985 to 5.446~eV, while their electron affinities (EA) range from 3.559 to 4.233~eV. HfPBr has the largest IP (5.446~eV), whereas TiPBr has the largest EA (4.233~eV). Bader charge analysis reveals positive charge-transfer values at the transition-metal sites and negative values at the pnictogen and halogen sites, indicating electron depletion at the metal sites and electron accumulation at the other constituent sites.

The calculated band gaps and vacuum-referenced band-edge positions provide a basis for evaluating the photocatalytic potential of these semiconductors through their alignment with the relevant electrochemical redox potentials. Semiconductor photocatalysis involves light-induced charge carriers that drive surface reduction and oxidation reactions. In water splitting, photogenerated electrons participate in the hydrogen evolution reaction (HER), \ce{2H+ + 2e- -> H2}, while holes drive the oxygen evolution reaction (OER), \ce{2H2O -> O2 + 4H+ + 4e-}. Alternatively, conduction-band electrons can facilitate \ce{CO2} reduction to products such as \ce{CO}, \ce{HCOOH}, \ce{CH3OH}, and \ce{CH4}, depending on the reaction pathway. The thermodynamic feasibility of these processes is governed by the relative positions of the conduction-band minimum (CBM), valence-band maximum (VBM), and the corresponding redox potentials.\cite{Nishioka2023} Consequently, evaluating band-edge alignment provides an initial criterion for screening semiconductors for potential photocatalytic applications, although it does not establish their catalytic activity.

The redox potentials of proton-coupled electron-transfer reactions depend on pH and temperature. For a redox couple involving equal numbers of protons and electrons, the Nernst relationship on the vacuum energy scale can be expressed as

\begin{equation}
\varepsilon_{\mathrm{redox}}(\mathrm{pH},T)
=
\varepsilon_{\mathrm{redox}}(0,T)
+
\frac{2.303 k_{\mathrm{B}}T}{e}\,\mathrm{pH}
\label{eq:nernst}
\end{equation}

where $\varepsilon_{\mathrm{redox}}(0,T)$ denotes the vacuum-referenced redox energy at $\mathrm{pH}=0$. At 25~$^\circ$C, the pH-dependent shift is approximately 0.0591~eV per pH unit. Accordingly, the conventional vacuum-referenced energy levels for HER ($\ce{H+/H2}$) and OER ($\ce{O2}/\ce{H2O}$) shift from approximately $-4.44$ and $-5.67$~eV at $\mathrm{pH}=0$ to $-4.03$ and $-5.26$~eV, respectively, at $\mathrm{pH}=7$.\cite{Wen2025,Ahmad2022} The minimum thermodynamic energy requirement for overall water splitting is 1.23~eV, which sets the fundamental energy threshold for the band gap of a semiconductor photocatalyst. However, satisfying this energy requirement alone is insufficient, as the band-edge positions must also straddle the HER and OER redox levels to enable overall water splitting. In practice, a larger band gap or additional driving force may be required to overcome reaction overpotentials and kinetic barriers.

For $\ce{CO2}$ reduction, the required reduction potential depends on the reaction pathway and target product. Based on the standard reduction potentials reported by Zhao et al.\cite{Zhao2020} at 25~$^\circ$C and $\mathrm{pH}=7$, the corresponding vacuum-referenced redox levels are approximately $-3.92$~eV ($-0.52$~V vs.~NHE) for $\ce{CO2}/\ce{CO}$, $-3.89$~eV ($-0.55$~V vs.~NHE) for $\ce{CO2}/\ce{HCOOH}$, $-4.05$~eV ($-0.39$~V vs.~NHE) for $\ce{CO2}/\ce{CH3OH}$, and $-4.19$~eV ($-0.25$~V vs.~NHE) for $\ce{CO2}/\ce{CH4}$. Comparing these levels with the calculated CBM provides an initial assessment of the thermodynamic feasibility of the corresponding reduction pathways. The VBM must also be suitably positioned to support the associated oxidation reaction.

The photocatalytic potential of the stable screened $\beta$-MPnX monolayers is evaluated based on these criteria, as illustrated in Fig.~\ref{fig:photocatalysis}. The calculated band-edge alignment identifies TiPI, ZrPBr, ZrAsBr, and HfPBr as candidates for overall water splitting at $\mathrm{pH}=7$. Their CBM positions lie above the HER redox level ($-4.03$~eV), while their VBM positions lie below the OER redox level ($-5.26$~eV), satisfying the thermodynamic requirements for both half-reactions. Their band gaps are 1.53, 1.55, 1.56, and 1.89~eV, respectively, all exceeding the minimum thermodynamic energy requirement of 1.23~eV.

For \ce{CO2} reduction, nine semiconductors satisfy the band-edge criterion for \ce{CO} formation, while eight satisfy that for \ce{HCOOH} formation. The criteria for \ce{CH3OH} and \ce{CH4} formation are met by all semiconductors except TiPBr. Notably, TiPI, ZrPBr, ZrAsBr, and HfPBr satisfy the band-edge criteria for both overall water splitting and all four considered \ce{CO2}-reduction pathways. This overlap identifies these four monolayers as the most versatile candidates in the present screening for further evaluation of water-splitting and \ce{CO2}-reduction photocatalysis.

\begin{table*}[!t]
\centering
\caption{Calculated effective masses ($m_x^\ast/m_0$ and $m_y^\ast/m_0$), in-plane elastic moduli ($C_{2D,x}$ and $C_{2D,y}$), deformation potentials ($E_{d,x}$ and $E_{d,y}$), coefficients of determination ($R_x^2$ and $R_y^2$), and carrier mobilities ($\mu_x$ and $\mu_y$) of the investigated monolayers. Elastic moduli are in $\mathrm{N\,m^{-1}}$, deformation potentials in eV, and mobilities in $\mathrm{cm^2\,V^{-1}\,s^{-1}}$. Mobilities are reported only when the corresponding deformation-potential fits satisfy $R^2 \geq 0.99$. ZrBiBr is excluded because it exhibits metallic behaviour.}
\label{tab:mobility}

\scriptsize
\setlength{\tabcolsep}{3pt}
\renewcommand{\arraystretch}{1.3}

\resizebox{\textwidth}{!}{%
\begin{tabular}{c c c c c c c c c c c c}
\hline
Compound & Carrier
& $m_x^\ast/m_0$ & $m_y^\ast/m_0$
& $C_{2D,x}$ & $C_{2D,y}$
& $E_{d,x}$ & $R_x^2$
& $E_{d,y}$ & $R_y^2$
& $\mu_x$ & $\mu_y$ \\
\hline

\multirow[c]{2}{*}{TiPBr}
& Electron & 0.84 & 0.75 & 69.45 & 70.17 & -2.05 & 0.99 & -0.57 & 0.98 & 530.45 & --- \\
& Hole     & 2.18 & 0.56 & 69.45 & 70.17 & -6.43 & 0.99 & -6.56 & 0.99 & 14.87 & 56.93 \\

\multirow[c]{2}{*}{TiPI}
& Electron & 1.05 & 1.04 & 65.87 & 65.84 & -3.41 & 0.99 & -1.83 & 0.99 & 109.45 & 384.64 \\
& Hole     & 1.22 & 2.60 & 65.87 & 65.84 & -7.73 & 0.99 & -6.17 & 0.99 & 10.78 & 7.96 \\

\multirow[c]{2}{*}{TiAsI}
& Electron & 2.43 & 2.67 & 53.98 & 54.37 & -2.46 & 0.92 & -2.47 & 0.92 & --- & --- \\
& Hole     & 1.08 & 0.84 & 53.98 & 54.37 & -3.35 & 0.99 & -3.03 & 0.99 & 99.19 & 157.93 \\

\multirow[c]{2}{*}{ZrPCl}
& Electron & 0.85 & 1.00 & 47.14 & 46.52 & -1.18 & 0.99 & 0.74 & 0.96 & 907.25 & --- \\
& Hole     & 0.60 & 0.79 & 47.14 & 46.52 & -8.90 & 0.99 & -6.04 & 0.99 & 30.93 & 50.38 \\

\multirow[c]{2}{*}{ZrPBr}
& Electron & 1.52 & 1.83 & 40.88 & 39.30 & -3.49 & 0.99 & -1.14 & 0.89 & 28.15 & --- \\
& Hole     & 1.34 & 0.96 & 40.88 & 39.30 & -7.63 & 0.99 & -6.59 & 0.99 & 9.82 & 17.67 \\

\multirow[c]{2}{*}{ZrPI}
& Electron & 0.97 & 0.93 & 20.95 & 20.57 & -3.83 & 0.93 & -3.81 & 0.92 & --- & --- \\
& Hole     & 1.46 & 1.46 & 20.95 & 20.57 & -3.98 & 0.99 & -3.95 & 0.99 & 13.24 & 13.27 \\

\multirow[c]{2}{*}{ZrAsBr}
& Electron & 2.78 & 3.02 & 32.06 & 33.23 & -2.54 & 0.82 & -2.57 & 0.82 & --- & --- \\
& Hole     & 1.08 & 0.92 & 32.06 & 33.23 & -7.00 & 0.99 & -4.70 & 0.99 & 12.95 & 35.00 \\

\multirow[c]{2}{*}{ZrAsI}
& Electron & 0.99 & 0.95 & 28.38 & 31.79 & -3.08 & 0.91 & -3.05 & 0.91 & --- & --- \\
& Hole     & 0.81 & 1.31 & 28.38 & 31.79 & -2.95 & 0.99 & -3.22 & 0.99 & 83.30 & 48.46 \\

\multirow[c]{2}{*}{HfPCl}
& Electron & 1.10 & 1.01 & 36.97 & 36.29 & -2.59 & 0.99 & -0.81 & 0.98 & 101.25 & --- \\
& Hole     & 0.88 & 0.59 & 36.97 & 36.29 & -7.56 & 0.99 & -6.99 & 0.99 & 21.66 & 37.20 \\

\multirow[c]{2}{*}{HfPBr}
& Electron & 1.85 & 2.04 & 33.49 & 33.87 & -4.12 & 0.99 & -1.93 & 0.98 & 11.74 & --- \\
& Hole     & 0.92 & 0.69 & 33.49 & 33.87 & -7.63 & 0.99 & -6.50 & 0.99 & 16.64 & 30.84 \\

\multirow[c]{2}{*}{HfAsI}
& Electron & 0.57 & 0.48 & 31.25 & 30.26 & -5.71 & 0.99 & -4.38 & 0.99 & 67.81 & 132.47 \\
& Hole     & 0.87 & 1.78 & 31.25 & 30.26 & -3.63 & 0.99 & -4.52 & 0.99 & 46.68 & 14.30 \\

\hline
\end{tabular}%
}
\end{table*}

Favourable band-edge alignment is a necessary but not sufficient condition for photocatalysis, because the reactions must also overcome kinetic barriers on the catalyst surface and compete with carrier recombination. The driving forces of the photogenerated carriers were estimated as the energy difference between the CBM and the HER level for electrons, and between the OER level and the VBM for holes. For the four candidates, the electron driving force ranges from 0.19 to 0.47~eV, whereas the hole driving force is only 0.06--0.19~eV. Because even the best oxide catalysts require a minimum OER overpotential of about 0.2--0.4~V,\cite{Man2011} hydrogen evolution is better supported than overall water splitting, and efficient OER would likely require co-catalysts or band-edge engineering. In addition, the halogen-terminated surfaces of $\beta$-MPnX are likely to bind reaction intermediates weakly, so edges or vacancies could serve as the active sites. The different character of the band edges could also help reduce recombination. Photogenerated electrons occupy conduction-band states dominated by transition-metal $d$ orbitals, whereas holes occupy valence-band states with substantial pnictogen $p$ contributions, which limits their spatial overlap.

\begin{figure}[!t]
\centering
\includegraphics[width=\columnwidth]{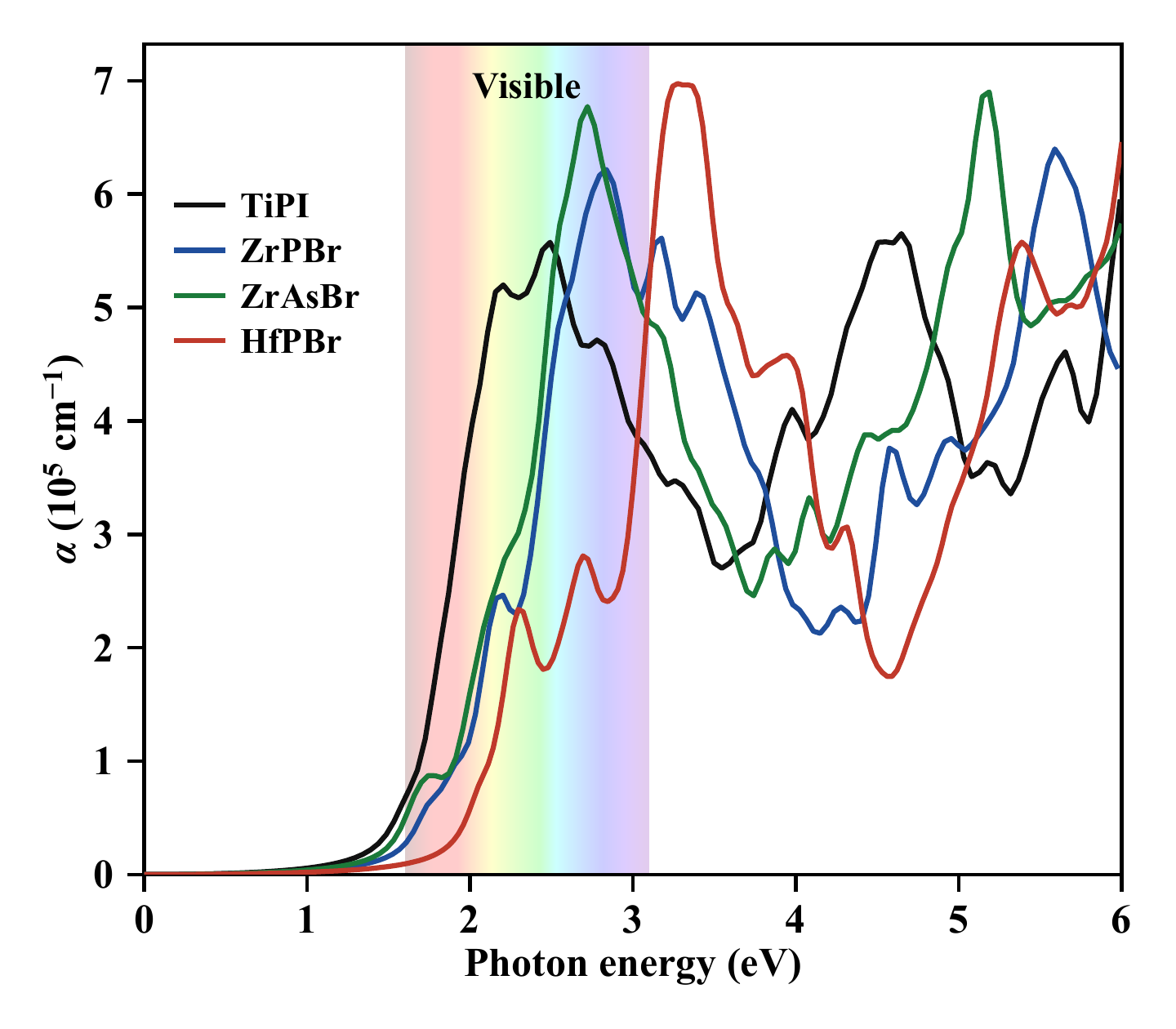}
\caption{Average in-plane optical absorption coefficients, $\alpha = (\alpha_{xx} + \alpha_{yy})/2$, of TiPI, ZrPBr, ZrAsBr and HfPBr calculated using HSE06. The shaded region marks the visible range (1.6--3.1~eV).}
\label{fig:absorption}
\end{figure}

Efficient photocatalysis also requires strong absorption of visible light to generate charge carriers. To assess the light-harvesting ability of the four candidates, their optical absorption spectra were calculated using HSE06, as shown in Fig.~\ref{fig:absorption}. The absorption is isotropic in the plane, and for each monolayer the onset coincides with its HSE06 band gap (Table~\ref{tab:electronic_properties}). Among them, TiPI absorbs most strongly at low photon energies, rising steeply from the red end of the visible range to a broad maximum at 2.2--2.5~eV. ZrAsBr and ZrPBr show their main absorption peaks in the blue region of the visible range, at about 2.7 and 2.8~eV, respectively. Owing to its larger band gap, HfPBr shows two weaker absorption peaks in the visible range, at about 2.3 and 2.7~eV, while its strongest absorption occurs just above the visible edge, at about 3.3~eV. All four monolayers thus absorb a substantial part of the visible spectrum, which, together with their band-edge alignment, supports their potential as visible-light photocatalysts. Because these spectra were calculated within the independent-particle approximation, they neglect electron--hole interactions. In two-dimensional semiconductors, the weak, non-local dielectric screening of the Coulomb interaction leads to strongly bound excitons,\cite{Chernikov2014} which shift the absorption onset below the quasiparticle gap and enhance the absorption near the band edge. The present spectra therefore provide a conservative estimate of the visible-light absorption. In photocatalysis, however, excitons must first dissociate into free electrons and holes before they can drive the surface redox reactions, so strong exciton binding may hinder charge separation.

\begin{figure*}[!t]
    \centering
    \includegraphics[width=0.90\textwidth]{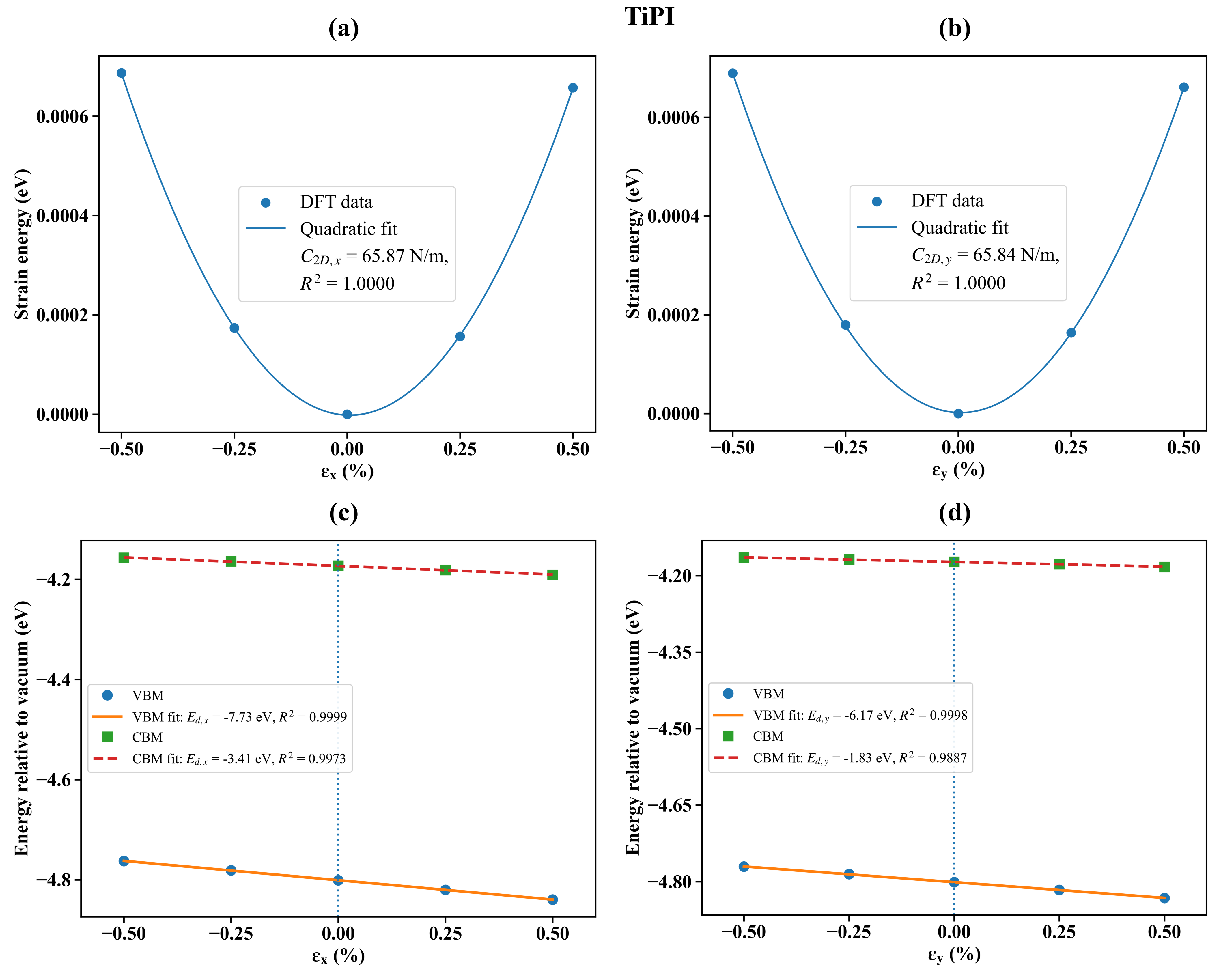}
    \caption{Strain-dependent total energy and band-edge positions of the TiPI monolayer under uniaxial strain along the in-plane $x$ and $y$-directions. (a,b) Strain--energy curves fitted using quadratic functions to determine the in-plane elastic moduli, $C_{2D,x}$ and $C_{2D,y}$, respectively. (c,d) Variations in the VBM and CBM energies relative to vacuum as functions of uniaxial strain along the $x$ and $y$-directions, respectively. Linear regression of the band-edge energies was used to determine the deformation potentials for holes and electrons, respectively. The corresponding coefficients of determination ($R^2$) indicate the quality of the fits.}
    \label{fig:mobility_tipi}
\end{figure*}

Beyond their potential for photocatalytic applications, the charge-transport properties of the stable monolayers are also important for assessing their broader functional potential. Carrier mobility provides a measure of how readily electrons and holes move through a material and can reveal the influence of transport anisotropy on their electronic performance. The anisotropic carrier mobility of the stable monolayers was evaluated using the Bardeen--Shockley deformation-potential theory.\cite{BardeenShockley1950} Within this framework, the carrier mobility along an in-plane direction $i$ is expressed as\cite{Mir2020,Lang2016}

\begin{equation}
\mu_i =
\frac{e\planckbar^3 C_{2D,i}}
{k_{\mathrm{B}}T\,m_i^\ast m_d E_{d,i}^{2}}
\label{eq:mobility}
\end{equation}
    
where $e$ is the elementary charge, $\planckbar$ is the reduced Planck constant, $k_\mathrm{B}$ is the Boltzmann constant, and $T$ is the absolute temperature. The parameters $C_{2D,i}$, $m_i^\ast$, $m_d$, and $E_{d,i}$ represent the in-plane elastic modulus, transport effective mass, density-of-states effective mass, and deformation potential, respectively. The transport effective mass was calculated from the curvature of the electronic band structure near the relevant band extremum,\cite{Spitzer1957}

\begin{equation}
m_i^\ast =
\frac{\planckbar^2}{\partial^2 E/\partial k_i^2}
\label{eq:effective_mass}
\end{equation}

where $E$ is the band energy and $k_i$ is the wave-vector component along the $i$-th direction ($i=x,y$). The two-dimensional density-of-states effective mass was calculated as $m_d = \sqrt{m_x^\ast m_y^\ast}$. The in-plane elastic modulus was obtained from the strain--energy response to uniaxial strain,\cite{Wang2022HighThroughput2D}

\begin{equation}
C_{2D,i} =
\frac{1}{A_0}
\frac{\partial^2 E_{\mathrm{strain}}}{\partial \varepsilon_i^2}
\label{eq:elastic_modulus}
\end{equation}

where $A_0$ is the equilibrium in-plane area, $E_{\mathrm{strain}}$ is the strain energy relative to the equilibrium structure, and $\varepsilon_i$ is the uniaxial strain along direction $i$. The deformation potential was determined from the slope of the band-edge energy with respect to strain,\cite{Spitzer1957}

\begin{equation}
E_{d,i} =
\frac{\partial E_{\mathrm{edge}}}{\partial \varepsilon_i}
\label{eq:deformation_potential}
\end{equation}

where $E_{\mathrm{edge}}$ denotes the conduction-band minimum (CBM) for electrons and the valence-band maximum (VBM) for holes.

The deformation potential strongly influences the calculated mobility because it appears squared in the denominator of Eq.~\ref{eq:mobility}. Consequently, deviations in the fitted band-edge energy--strain slope can substantially affect the resulting mobility. The linearity of each band-edge energy--strain fit was therefore assessed using the coefficient of determination ($R^2$). A threshold of $R^2\geq0.99$ was applied independently to the CBM and VBM fits along both in-plane directions. Only mobilities associated with fits meeting this criterion were retained for quantitative comparison, while the remaining values were excluded from the reported results.

Fig.~\ref{fig:mobility_tipi} presents TiPI as a representative example of the strain--energy and band-edge analyses used to determine the in-plane elastic moduli and deformation potentials. The corresponding results for the other investigated monolayers are provided in Figs. S23–S32. The calculated effective masses, elastic moduli, deformation potentials, fitting coefficients ($R^2$), and carrier mobilities are summarized in Table~\ref{tab:mobility}.

The calculated carrier mobilities reveal substantial differences among the investigated monolayers and between the two in-plane transport directions. For electron transport, ZrPCl exhibits the highest mobility along the $x$-direction, reaching $907.25~\mathrm{cm^2\,V^{-1}\,s^{-1}}$, followed by TiPBr at $530.45~\mathrm{cm^2\,V^{-1}\,s^{-1}}$. TiPI exhibits a high electron mobility of $384.64~\mathrm{cm^2\,V^{-1}\,s^{-1}}$ along the $y$-direction, whereas its mobility along the $x$-direction is $109.45~\mathrm{cm^2\,V^{-1}\,s^{-1}}$. HfPCl also exhibits an appreciable electron mobility of $101.25~\mathrm{cm^2\,V^{-1}\,s^{-1}}$ along the $x$-direction. For hole transport, TiAsI shows mobilities of $99.19$ and $157.93~\mathrm{cm^2\,V^{-1}\,s^{-1}}$ along the $x$ and $y$-directions, respectively. ZrAsI also exhibits relatively high hole mobilities of $83.30$ and $48.46~\mathrm{cm^2\,V^{-1}\,s^{-1}}$ along these directions. Overall, these results highlight the potential of ZrPCl and TiPBr for electron transport, TiPI for direction-dependent electron transport, and TiAsI for hole transport. These trends follow from the metal-$d$ character of the conduction-band edge. Replacing Cl with Br, or Zr with Hf, increases both the electron effective mass and the deformation potential. From HfPCl to HfPBr, for example, $m_x^\ast$ increases from 1.10 to 1.85~$m_0$ and $E_{d,x}$ from 2.59 to 4.12~eV, which lowers the electron mobility from 101.25 to 11.74~$\mathrm{cm^2\,V^{-1}\,s^{-1}}$. Similar reductions occur from ZrPCl to ZrPBr and from ZrPCl to HfPCl. The heavier masses indicate flatter conduction bands, and the larger deformation potentials indicate stronger coupling of the band edge to strain.

These mobilities should be interpreted within the limits of the deformation-potential approach, which treats only longitudinal acoustic-phonon scattering and neglects optical-phonon, intervalley and impurity scattering. Optical-phonon scattering is likely to be significant here, because the Bader analysis shows substantial charge transfer from the metal to the pnictogen and halogen atoms, which indicates a polar lattice. Optical-phonon scattering has also been shown to reduce the mobility of the related $\beta$-ZrNCl and $\beta$-HfNCl monolayers.\cite{Liu2025} The calculated values are therefore upper estimates. The high electron mobility of ZrPCl arises mainly from its small deformation potential of 1.18~eV along $x$, compared with 1.83--5.71~eV for the other monolayers with reported electron mobilities, which indicates weaker acoustic-phonon scattering. As a result, its real mobility is likely limited mainly by optical-phonon scattering, which is not included here. The calculated value for ZrPCl may thus exceed the real mobility by a larger margin than for the other monolayers, whereas the trends across compositions and directions are more reliable.

\section*{Conclusions}
The combined stability assessments reveal that, despite favourable formation and cohesive energies across all 48 $\beta$-MPnX monolayers, only 12 satisfy the energetic, mechanical, and dynamical stability criteria. Notably, 25 of the 26 metallic compounds exhibit phonon instabilities. \textit{Ab initio} molecular dynamics simulations at 300~K further reveal that 10 of the 12 selected monolayers retain their structural integrity over 5~ps, whereas ZrAsBr and HfPCl exhibit finite-temperature structural instability. Extended 10~ps simulations confirm the stability of TiPI, ZrPBr and HfPBr.

HSE06 calculations predict semiconducting behaviour for 11 of the 12 selected monolayers, with band gaps ranging from 0.895~eV for TiPBr to 1.887~eV for HfPBr, whereas ZrBiBr is metallic. Bader charge analysis indicates charge transfer from the transition-metal atoms to the pnictogen and halogen atoms. The calculated work functions of the semiconducting monolayers range from 4.741~eV for ZrPCl to 5.250~eV for HfPBr, while their ionization potentials and electron affinities range from 4.985 to 5.446~eV and from 3.559 to 4.233~eV, respectively. Band-edge alignment identifies TiPI, ZrPBr, ZrAsBr, and HfPBr as promising candidates that satisfy the calculated thermodynamic requirements for overall water splitting and the four investigated \ce{CO2}-reduction pathways, corresponding to the formation of \ce{CO}, \ce{HCOOH}, \ce{CH3OH}, and \ce{CH4}. The calculated driving forces indicate that hydrogen evolution is better supported than oxygen evolution. The four candidates also show strong absorption across the visible range.

Carrier-transport calculations reveal substantial compositional and directional dependence. ZrPCl exhibits the highest reported electron mobility, reaching $907.25~\mathrm{cm^2\,V^{-1}\,s^{-1}}$ along the $x$-direction, followed by TiPBr, with $530.45~\mathrm{cm^2\,V^{-1}\,s^{-1}}$ along the same direction. For hole transport, TiAsI exhibits mobilities of 99.19 and $157.93~\mathrm{cm^2\,V^{-1}\,s^{-1}}$
along the $x$ and $y$-directions, respectively.

Collectively, these findings establish a systematic framework for investigating the relationships between composition, stability, electronic properties, photocatalytic band-edge alignment, and carrier transport in $\beta$-MPnX monolayers. This work expands the computationally explored landscape of two-dimensional transition-metal pnictide halides and provides a foundation for prioritizing promising candidates for future theoretical and experimental investigations of photocatalytic and charge-transport properties.

\section*{Author contributions}

M. T. A.: conceptualization, data curation, formal analysis, investigation, software, visualization, writing – original draft. S. H. N.: supervision, writing – review \& editing. M. S. A.: supervision, project administration, writing – review \& editing.

\section*{Declaration of competing interest}

The authors declare that they have no known competing financial interests or personal relationships that could have appeared to influence the work reported in this paper.

\section*{Data availability}

The data that support the findings of this study are available from the corresponding author upon reasonable request, and additional data are included in the Supplementary Material, provided as a separate file (Tables S1--S2 and Figs. S1--S32).

\section*{Acknowledgements}

The authors gratefully acknowledge Pabna University of Science and Technology (PUST) for institutional support and the TEAMS Lab at PUST for providing computational facilities.

%%% END OF MAIN TEXT %%%

%%% REFERENCES %%%
\bibliography{references}
\bibliographystyle{numbered}

\end{document}